# Ferroelectric control of the giant Rashba spin orbit coupling in GeTe(111)/InP(111) superlattice

Yu-Hua Meng[1], Wei Bai[1], Heng Gao[2], Shi-Jing Gong[1*], Chun-Gang Duan[1,3], Jun-Hao Chu[1]

[1]*Key Laboratory of Polar Materials and Devices, Ministry of Education, East China Normal University, Shanghai 200062, China*

[2]*Department of Physics, Shanghai University, Shanghai 200444, China*

[3]*Collaborative Innovation Center of Extreme Optics, Shanxi University, Taiyuan, Shanxi 030006, China*

**E-mail:** *sjgong@ee.ecnu.edu.cn*

**Abstract**

GeTe wins the renewed research interest due to its giant bulk Rashba spin orbit coupling (SOC), and becomes the father of a new multifunctional material, *i.e.*, ferroelectric Rashba semiconductor. In the present work, we investigate Rashba SOC at the interface of the ferroelectric semiconductor superlattice GeTe(111)/InP(111) by using the first principles calculation. Contribution of the interface electric field and the ferroelectric field to Rashba SOC is revealed. A large modulation to Rashba SOC and a reversal of the spin polarization is obtained by switching the ferroelectric polarization. Our investigation about GeTe(111)/InP(111) superlattice is of great importance in the application of ferroelectric Rashba semiconductor in the spin field effect transistor.

Spintronics deals with the manipulation of the electron's spin degree of freedom. Within the field of spintronics, much attention has been focused on Rashba spin-orbit coupling (SOC),[1, 2] which arises from the structure inversion asymmetry (SIA) at the surface/interface systems, and provides an effective in-plane k-dependent magnetic field for the electrons moving in the two dimensional surface/interface systems. More importantly, the gate tunability of Rashba SOC enables an all-electric way of controlling the spin degree of freedom,[3-9] such as the Datta-Das spin field effect transistor (FET)[10], in which the spin procession in the InGaAs/InAlAs 2DEG can be precisely manipulated by Rashba SOC. More recently, an all-electric and all-semiconductor spin FET based on the modulation of Rashbba SOC has been experimentally realized.[4] Up till now, the tunability of Rashba SOC strength is still one of the most important issues in the field of spintronics, and has been widely investigated in various systems, such as the semiconductor two dimensional electron gas (2DEG) systems,[11-16] the metal surfaces, [17-19] etc.

Since the effective electric field induced by the polarization charge at the interface can be much higher than the electric field produced in the laboratory conditions, Rashba SOC in polar materials is expected to be much larger, which has been proved in the polar layer compound BiTeX (X = Br and I) [20] and TMD monolayers MXY (M=Mo, W, X≠Y=S, Se, Te).[21] Ferroelectrics is a big family with the spontaneous polarization, and ferroelectric polarization can be easily reversed at relatively low voltages, which means that Rashba SOC can be easily modulated by a small electric field and Rashba SOC resulting from the ferroelectric field can take on the bistability and the nonvolatility of the ferroelectrics. In the Bi/BaTiO$_3$ interface, Mirhosseini and Abdelouahed *et al* has demonstrated that, the Rashba spin splitting in the 6p states of a Bi adlayer on BaTiO3(001) can be manipulated by the electric polarization in the ferroelectric substrate.[22, 23] For a long time, the interest in Rashba SOC has been focused on the surface/interface system, until recently, Rashba SOC in polar bulk materials has been reported.[24-26] In ferroelectric semiconductor GeTe, strong bulk Rashba SOC and a large tunability were reported by Picozzi *et al*,[9] based on which a new class of multifunctional material named as ferroelectric Rashba semiconductor (FRS)[27, 28] is proposed, the ferroelectric controlled spin FET has been proposed.

Inspired by the growing research interest in GeTe[29-36] and the significance of the ferroelectric controlled spin FET, we presently investigate Rashba SOC in the ferroelectric superlattice GeTe(111)/InP(111). The manipulation of Rashba SOC at the superlattice interface are discussed. Giant Rashba SOC and large tunability can be obtained through ferroelectric modulation. The evolution of the spin polarization are also investigated, and the flexible switching of up-spin, down-spin, and unpolarized spin can be realized.

First-principles calculations based on the density-function theory (DFT) are performed to explore the atomic and band structure of GeTe(111)/InP(111) superlattice by using the projector augmented wave (PAW) method implemented in Vienna Ab-initio Simulation Package (VASP).[37] The SOC is included in the calculations. The exchange-correlation potential is treated in the Perdew-Burke-Ernzerhof (PBE) with the energy cutoff of 500eV for the plane wave expansion and a 11*11*1 k-point gird are used in the self-consistent calculations.[38] The convergence of the total energy has been checked by changing the number of sampling k-points, energy cutoff, and the thickness of vacuum space. The structures are relaxed until the Hellmann-Feynman forces on each atom are less than 1 meV/Å.

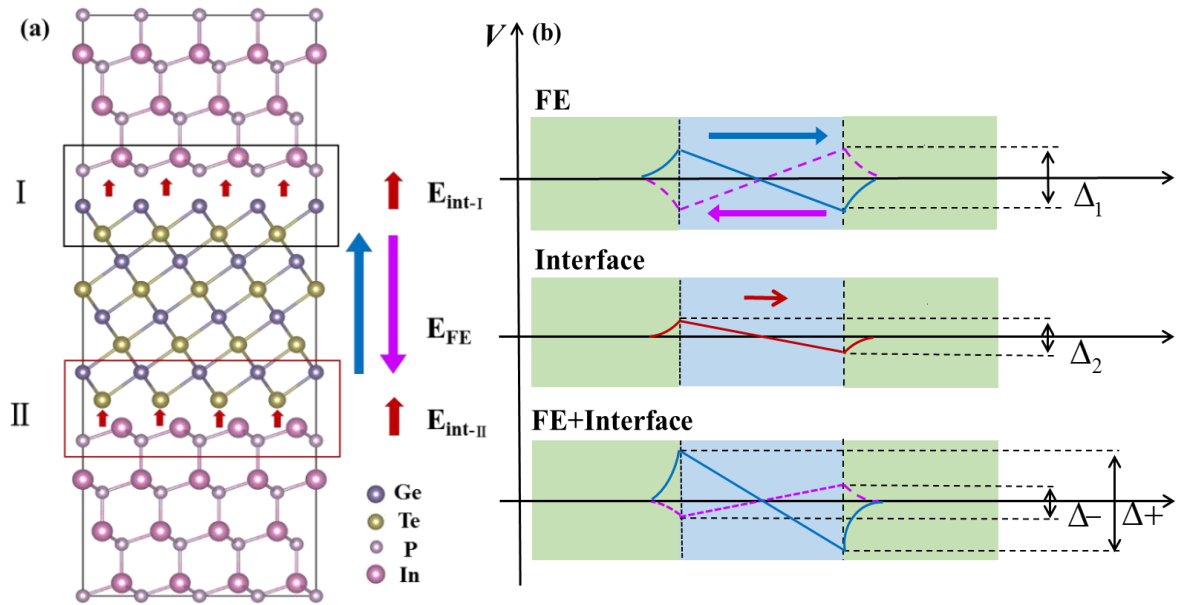

FIG. 1. (a) The atomic structure of GeTe(111)/InP(111) supercell, in which four layers of GeTe are sandwiched between the InP layers, the interface electric fields at the two different interfaces are indicated by the red arrows, and the ferroelectric field are shown in blue and purple arrows. (b) Schematic illustration of the electrostatic potential through the supercell due to the contribution of the bistable ferroelectric field (top), the asymmetric interface field (middle), and the combination of both the ferroelectric and interface field (bottom).

Shown in Fig. 1(a) is the atomic structure of GeTe(111)/InP(111) superlattice, in which four layers of GeTe are sandwiched between InP layers, and Ge, Te, In, and P atoms are represented by the solid circles in red, grey, blue and yellow, respectively. Bulk InP has the zinc-blend structure with the lattice constant 4.37 Å and bulk GeTe has the rhombohedral crystal structure with the lattice constant 5.87 Å, which means a good match between GeTe(111) and InP(111) ( a mismatch only about 1.7%). There are two interfaces in the superlattice, one with Te-In bonding and the other with Ge-P bonding, which are labelled by I and II in Fig. 1(a). The interface electric fields at the two interfaces are denoted by $E_{int}$-I and $E_{int}$-II, respectively. In addition, there also exists an electric field induced by the ferroelectric polarization $E_{FE}$, which can be positive or negative. Here, the positive ferroelectric field has same the direction as the interface electric field. The positive and negative ferroelectric fields are shown in blue and purple, respectively. The distributions of the interface potential are schematically shown in Fig. 1(b). The ferroelectric potential with the maximal magnitude $\Delta_1$ and the interface potential with the maximal magnitude $\Delta_2$ are individually displayed in the top and middle panes of Fig. 1(b). In the bottom pane of Fig. 1(b), they are combined to form the genuine potential at the interface, in which $\Delta_+$ indicates the ferroelectric and interface field have the same direction, and $\Delta_-$ means the ferroelectric field is partly cancelled by the interface electric field.

To quantitatively show the dependence of the bulk Rashba SOC on the strength of the ferroelectric polarization, we calculate the Rashba splitting bands under different ferroelectric displacements in bulk GeTe(111). The relative shift of the Ge and Te sublattices is reported using the distorted rocksalt setting: atomic positions are

(0, 0, 0) and (0.5-τ, 0.5-τ, 0.5-τ) with τ=0.03 for Ge and Te, respectively. Fig. 2(a) shows the spin splitting band structure of the bulk GeTe(111), and the inset is the two-dimensional plane of the first brillouin zone, in which the high symmetry k-points K, K′, M, Γ, and the reciprocal lattice constants are indicated. Fig. 2(b) shows the Rashba spin splitting energy under different ferroelectric displacements. We find that Rashba splitting energy is linearly dependent on the wavevectors around the Γ point. With the ferroelectric displacement τ, we find Rashba spltting energy can be up to 300 meV, and Rashba parameter can reach 6.3 eV·Å. Compare the different curves shown in Fig. 2(b), we notice that Rashba splitting energy is not linearly dependent on the ferroelectric displacements. When the ferroelectric displacement is τ/2, Rashba parameter is 4.4 eV·Å, which means that the giant Rashba SOC survives at room temperature.

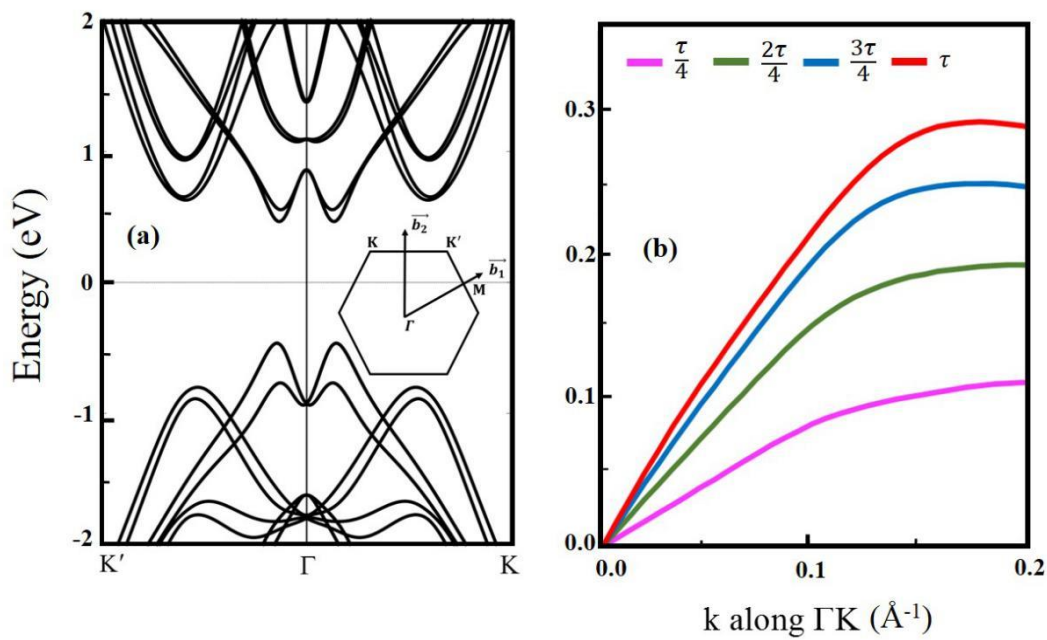

FIG. 2. (Color online) (a) Band structure of the bulk GeTe(111) along the high symmetry k-points K′-Γ-K, and the inset is the brillouin zone of the GeTe(111) surface. (b) Rashba spin splitting under different ferroelectric displacements.

Band structures of the GeTe(111)/InP(111) superlattice are shown in Fig. 3, in which the radius of the green, blue and pink circles indicate the weight of the $p_x$, $p_y$, $p_z$ orbital of Te atom at the interface-I shown in Fig. 1(a). The orbital analysis shows that the Rashba splitting bands are mainly contributed by the $p_z$ orbital of Te atoms, which is more likely be influenced by the z-oriented electric field at the interface. To reveal the influence of the ferroelectric field on Rashba SOC, we consider the paraelectric and ferroelectric phase for GeTe. Band structure for the paraelectric GeTe is shown in Fig. 3(a), in which we can see that, without the ferroelectric field, the interface electric field can still induce a minor Rashba spin splitting around the Γ point. We then calculate the band structures with the positive and negative ferroelectric fields, respectively. It can be clearly seen that the superlattice shows the largest Rashba SOC when the ferroelectric field is parallel to the interface electric field, as shown in Fig. 3 (b). When the ferroelectric field is antiparallel to the interface electric field, Rashba SOC strength is larger than the paraelectric phase while smaller than the positive ferroelectric phase, which can be seen by comparing Fig. 3(a)-(c).

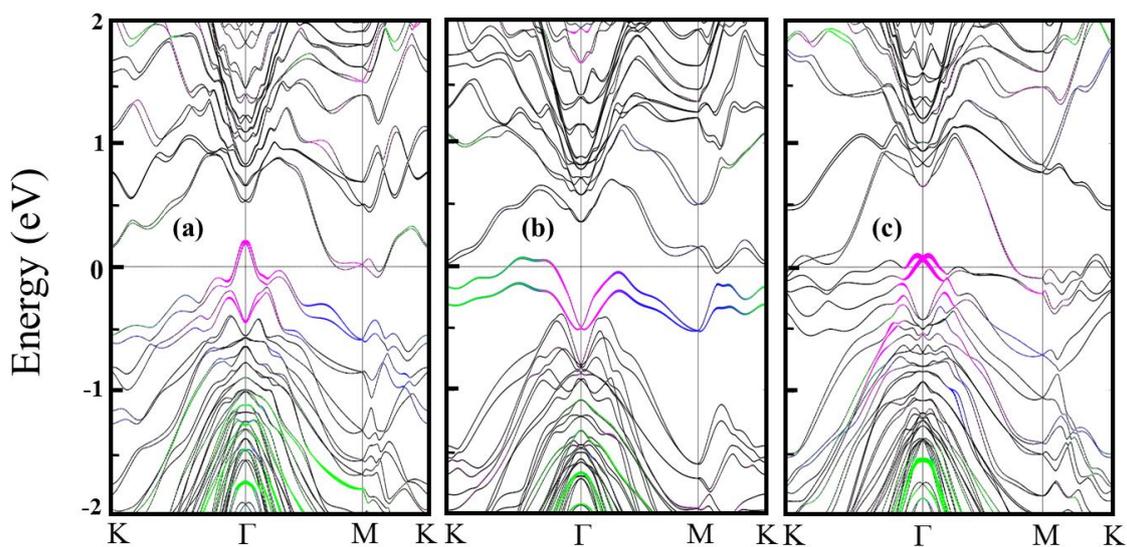

FIG. 3. The orbital-projected band structures of GeTe(111)/InP(111) superlattice, with the radius of the green, blue and pink circles indicate the weight of the $p_x$, $p_y$, $p_z$ orbital of Te atom at the interface-I shown in Fig.1(a). (a) GeTe is in the paraelectric phase, (b) and (c) GeTe is in the ferroelectric phase with upward and downward polarization, respectively. The green, blue and purple circles indicate the $p_x$, $p_y$, $p_z$ orbital of Te atom at the interface.

To well understand the involution of Rashba SOC induced by the change of the ferroelectric field, we compare Rashba splitting bands under different ferroelectric placements. In Fig. 4(a), we show the results for the positive ferroelectric field, with the ferroelectric displacements 0, +τ/3, +2τ/3, and +τ. Since the positive ferroelectric field has the same direction as the interface electric field, the increase of the ferroelectric displacement will surely result in a constant increase of the Rashba SOC, which is accompanied with the gradual shifting down of the energy levels of the Rashba splitting bands. On the other hand, when the ferroelectric displacement is netaive, i.e., the ferroelectric field is antiparallel to the interface electric field, Rashba SOC decreases to zero when the ferroelectric displacement is about -τ/4, which means that the ferroelectric field is equal to the interface electric field. In Fig. 4(b), we show the results for the negative ferroelectric displacements 0, -τ/3, -2τ/3, -τ. With the ferroelectric displacement changing from 0 to -τ, Rashba SOC decreases to zero at -τ/4 (not shown) and then recover. Also, energy levels of the Rashba splitting bands shifts upward from 0 to -τ/4 (not shown) and then downward from -τ/4 to -τ.

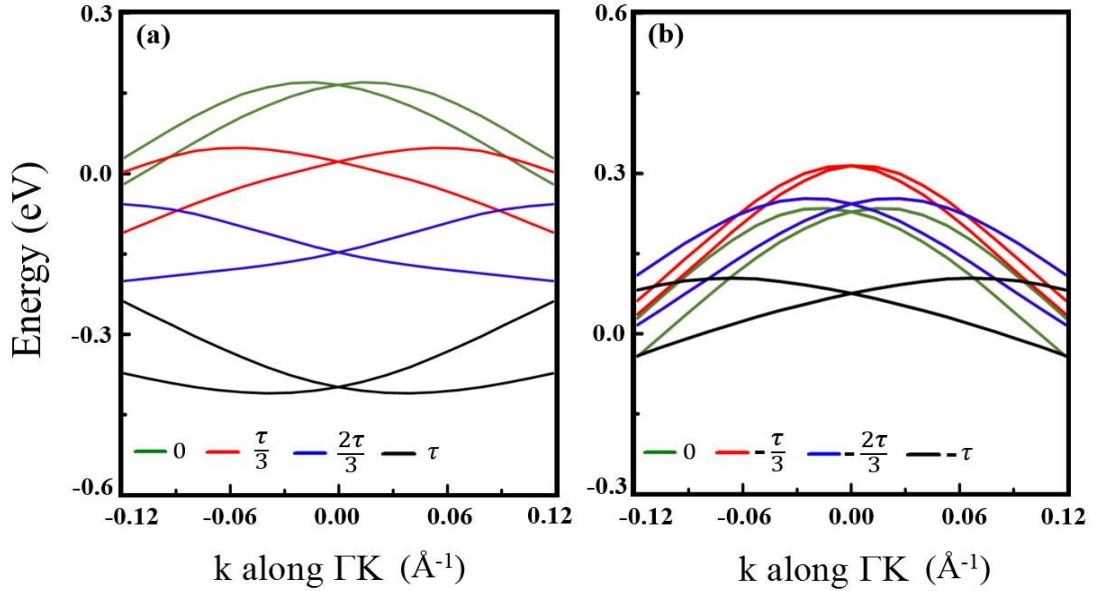

FIG. 4. Rashba splitting bands under (a) positive and (b) negative ferroelectric displacements.

We finally plot the schematic diagram of Rashba SOC versus the ferroelectric displacements in Fig. 5, with the two insets representing the two spin distribution under different ferroelectric displacements. The arrows indicate the magnitude of the in-plane spin polarization, and the red/blue color indicate the out-of-plane spin polarization. The dotted line is located at the critical ferroelectric displacement -τ/4. It can be clearly seen that on the right side of the critical displacement, the spin polarization is distributed anticlockwise, while on the left side of the critical field, the spin is clockwise. It is the direction of the interface electric field that determines the spin distribution. In addition, we also notice that from the Γ point to the K(K′) point, the spin polarization turns from the in-plane to the out-of-plane, and the spin polarization at K and K′ has opposite directions, which can possibly applied in the valleytronics.

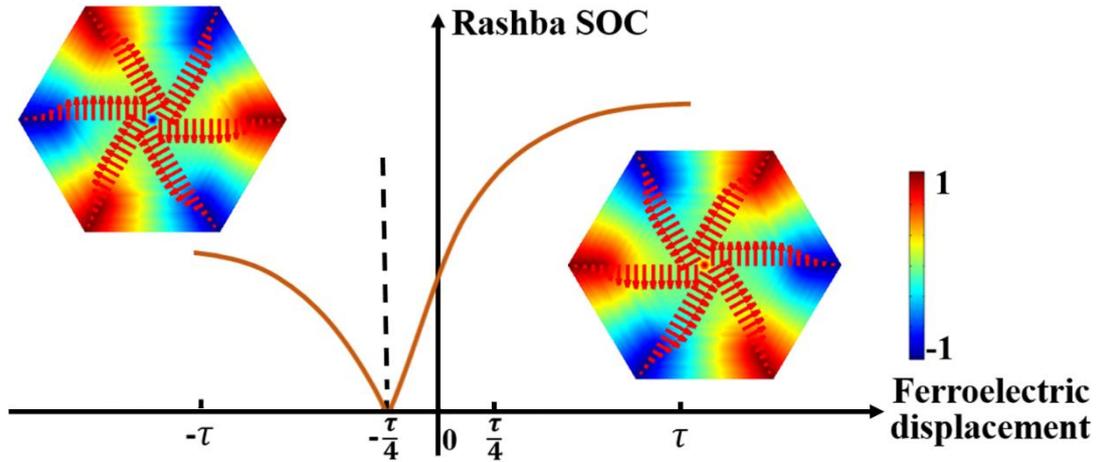

FIG. 5. The schematic diagram of Rashba SOC versus the ferroelectric displacements, with the dashed line indicating the critical ferroelectric displacement at which Rashba SOC is zero. At the two sides of the critical displacement, the two contours show the spin distribution in the topmost valence band along the line ΓK(K′), with the red arrows indicating the spin polarization in the *x-y* plane, and blue/red contour representing the spin polarization along the z direction.

**Conclusion**

By using the first-principles calculations, we investigate Rashba SOC in the ferroelectric semiconductor superlattice GeTe(111)/InP(111). A giant interface Rashba SOC is obtained when the ferroelectric field has the same direction as the interface field, and a large modulation of Rashba SOC is easily obtained due to the tunability of the ferroelectricity by a relatively weak electric field. We further find that the spin distribution for the topmost valence band can be reversed 180º by switching the ferroelectricity, in which Rashba SOC works as a bridge that connects the spin and charge degrees of freedom in the ferroelectric superlattice. Our investigation about GeTe(111)/InP(111) superlattice can help the application of ferroelectric Rashba semiconductor in the spin field effect transistor.

## Acknowledgement

This work was supported by the National Key Project for Basic Research of China (Grant Nos. 2014CB921104 and 2013CB922301), the National Natural Science Foundation of China (Grant No. 51572085 and 61674058). Computations were performed at the ECNU computing center.